\documentstyle[12pt]{article}

\begin{document}

\begin{center}
{\Large{\bf Formation of Directed Beams from Atom Lasers} \\ [5mm]
V. I. Yukalov and E.P. Yukalova} \\ [3mm]

{\it Instituto de Fisica de S\~ao Carlos, Universidade de S\~ao Paulo\\
Caixa Postal 369, S\~ao Carlos, S\~ao Paulo 13560--970, Brazil\\
and \\
Bogolubov Laboratory of Theoretical Physics \\
Joint Institute for Nuclear Research, Dubna 141980, Russia}

\end{center}

\vskip 2cm

\begin{abstract}

The problem of creating well--collimated beams of atoms escaping from a
trap is studied. This problem is of high importance for the realization
of atom lasers. Nonadiabatic dynamics of neutral atoms in nonuniform
magnetic fields, typical of quadrupole magnetic traps, has been considered.
The main result of this report is that an unusual semiconfining regime of 
motion is found, when atoms are confined from one side of an axis but 
are not confined from another side. This regime can be achieved by means
of only magnetic fields. The formed atomic beam can be forwarded in
arbitrary direction.

\end{abstract}

\newpage

\section{Introduction}

The possibility of realizing Bose--Einstein condensation in trapped dilute
gases (see reviews [1,2]) demonstrated that a macroscopic number of bosons
could be produced in a single quantum state of trapped atoms. The occupation
of a single quantum state by a large number of bosons is the matter--wave
analog of the storage of photons in a single mode of a laser cavity. The
device that could emit coherent beams of Bose atoms, similarly to the
emission of photon rays by light lasers, has been called {\it atom laser}
[3--10]. Output couplers that are used for extracting condensed atoms from
a trap are based on transferring atoms from a magnetically trapped state
to another internal state that is not trapped [11--14]. The transferring
field can be either a weak radiofrequency field including spin flips between
trapped and untrapped states [11--14] or a field of a laser beam stimulating
Raman transitions between magnetic sublevels [15]. In the first case, atoms
in the output state simply fall down from the trap under the action of
gravity. In the second case, when the atoms are transferred to a magnetic
untrapped state, they are also given a momentum kick from the photon recoil,
so that they exit the trap in a well--defined beam whose direction and
velocity are governed by the irradiating Raman laser.

Since the very first condition on any laser is that its output be highly
directional [9], it is important to develop mechanisms permitting to create
well--collimated beams of atoms coupled out of a trap. A novel mechanism of
creating highly directional beams if atoms leaving a trap has recently been
suggested [16] and different regimes of operations have been studied [17--20].

This mechanism is based on transferring trapped atoms to a nonadiabatic
initial state, which can be done by a short triggering pulse. There exist
such initial spin states for which the motion of atoms becomes
{\it semiconfined} for the same configuration of the trap magnetic field.
Then atoms quit the trap flying out predominantly in one direction and
forming a well--collimated beam. In this report, we give a brief outline of
the semiconfining regime of motion.

\section{Solution of Differential Equations}

The main equations of motion for an atom of mass $m$ and magnetic moment 
$\mu_0$ are usually written in the semiclassical approximation for 
the quantum--mechanical average of the real--space coordinate, 
$\stackrel{\rightarrow}{R}=\{ R_\alpha\}$, and for the average 
$\stackrel{\rightarrow}{S}=\{ S_\alpha\}$ of the spin operator, with 
$\alpha=x,y,z$. The first equation writes
\begin{equation}
\label{1}
\frac{d^2R_\alpha}{dt^2} = \frac{\mu_0}{m}\stackrel{\rightarrow}{S}
\cdot \frac{\partial\stackrel{\rightarrow}{B}}{\partial R_\alpha} \; ,
\end{equation}
where $\stackrel{\rightarrow}{B}$ is a magnetic field and, for a while, 
we omit the gravitational force and a collision term whose role will be 
discussed later on. The equation for the average spin is
\begin{equation}
\label{2}
\frac{d\stackrel{\rightarrow}{S}}{dt} = 
\frac{\mu_0}{\hbar}\stackrel{\rightarrow}{S}
\times \stackrel{\rightarrow}{B} \; .
\end{equation}
The total magnetic field $\stackrel{\rightarrow}{B}=
\stackrel{\rightarrow}{B}_1 +\stackrel{\rightarrow}{B}_2$ consists of two 
terms. The first is the quadrupole field
\begin{equation}
\label{3}
\stackrel{\rightarrow}{B}_1 = B_1'\left ( R_x\stackrel{\rightarrow}{e}_x
+ R_y\stackrel{\rightarrow}{e}_y + \lambda R_z\stackrel{\rightarrow}{e}_z
\right ) \; , 
 \end{equation}
which we write in a slightly more general form than in Ref. [16], including 
the anisotropy parameter $\lambda$. The latter was taken in Ref. [16] in 
the standard form as $\lambda=-2$. But, if several magnetic coils are 
involved in the formation of the field (3), the anisotropy parameter 
$\lambda$ can be varied. The possibility of such a variation will be 
important in what follows. The second part of the magnetic field 
$\stackrel{\rightarrow}{B}$ is a transverse field
\begin{equation}
\label{4}
\stackrel{\rightarrow}{B}_2 = B_2\stackrel{\rightarrow}{h}(t) \; ,
\qquad \stackrel{\rightarrow}{h}(t) = h_x\stackrel{\rightarrow}{e}_x + 
h_y\stackrel{\rightarrow}{e}_y \; , 
\end{equation}
in which $h_\alpha=h_\alpha(t)$ and $|\stackrel{\rightarrow}{h}(t)|=1$. 
In Ref. [16] the rotating transverse field, with
\begin{equation}
\label{5}
h_x=\cos\omega t \; , \qquad h_y =\sin\omega t \; ,
\end{equation}
was considered, although, as will be shown here, this is not the sole 
possibility. Introducing the dimensionless space variable 
$\stackrel{\rightarrow}{r}\equiv\stackrel{\rightarrow}{R}/R_0 =\{ x,y,z\}$, 
measured in units of the length $R_0\equiv B_2/B_1'$, and defining the 
characteristic frequencies 
\begin{equation}
\label{6}
\omega_1\equiv \sqrt{\frac{\mu_0B_1'}{mR_0}} \; ' \qquad 
\omega_2 \equiv \frac{\mu_0 B_2}{\hbar} \; ,
\end{equation}
we may rewrite Eq. (1) as
\begin{equation}
\label{7}
\frac{d^2\stackrel{\rightarrow}{r}}{dt^2} = \omega_1^2
\left ( S_x \stackrel{\rightarrow}{e}_x + 
S_y \stackrel{\rightarrow}{e}_y + 
\lambda S_z \stackrel{\rightarrow}{e}_z\right ) 
\end{equation}
and the spin equation (2) as
\begin{equation}
\label{8}
\frac{d\stackrel{\rightarrow}{S}}{dt} =\omega_2\hat A
\stackrel{\rightarrow}{S} \; ,
\end{equation}
where the matrix $\hat A=[A_{\alpha\beta}]$ is antisymmetric, 
$A_{\alpha\beta}= - A_{\beta\alpha},\; A_{\alpha\alpha}=0$, and
\begin{equation}
\label{9}
A_{12}= \lambda z \; , \qquad A_{23} = x +h_x \; , 
\qquad A_{31} = y + h_y \; .
\end{equation}
Keeping in mind the following discussion on the generality of the 
approach, we do not require here that $h_x$ and $h_y$ in Eq. (9) be 
necessarily of the rotating form (5). What is required is the existence 
of {\it small parameters}, one of which is
\begin{equation}
\label{10}
\frac{\omega_1}{\omega_2} \ll 1
\end{equation}
and another one is
\begin{equation}
\label{11}
\frac{\omega}{\omega_2} \ll 1 \; , \qquad
\omega\equiv \max_t \left | \frac{d\stackrel{\rightarrow}{h}}{dt}
\right | \; .
\end{equation}
In the particular case of the rotating field (5), the value of $\omega$ 
in Eq. (11) coincides with the rotation frequency.

It is worth noting that the inequalities (10) and (11) are assumed from the
very beginning. Numerical estimates of Ref. [16] give $\omega_1/\omega_2\sim
10^{-5}$ and $\omega/\omega_2\sim 10^{-3}$. The existence of the small 
parameters (10) and (11) allows us to apply for solving the system of 
equations (7) and (8) averaging methods [21--23] as developed for 
multifrequency systems [24,25]. A generalization of averaging methods, 
called the scale separation approach [26], has been developed for 
treating statistical systems that can be described by stochastic 
differential equations. However, while equations (7) and (8) contain no 
stochastic terms, the mathematical foundation of solving them is directly 
based on the averaging methods [21--25]. 

According to the small parameters (10) and (11), the functional variables
$\stackrel{\rightarrow}{r}$ and $\stackrel{\rightarrow}{h}$ are to be 
treated as slow compared to the fast variable $\stackrel{\rightarrow}{S}$.
This means that $\stackrel{\rightarrow}{r}$ and $\stackrel{\rightarrow}{h}$
change very little, being almost constant, during the oscillation time of
$\stackrel{\rightarrow}{S}$. Therefore, $\stackrel{\rightarrow}{r}$ and
$\stackrel{\rightarrow}{h}$ can be regarded as quasi--invariants with respect
to $\stackrel{\rightarrow}{S}$. Such quasi--invariants in the theory of 
classical Hamiltonian systems are called adiabatic invariants [25], whose 
existence should not be confused with the adiabatic approximation.
Following the averaging methods [21--25], we can solve equation (8) for the 
fast spin variable with the slow variables $\stackrel{\rightarrow}{r}$
and $\stackrel{\rightarrow}{h}$ treated as quasi--invariants. The 
difference between the latter is that $\stackrel{\rightarrow}{r}$, as a 
function of time, is defined implicitly by Eq. (7), while 
$\stackrel{\rightarrow}{h}$ is an explicitly given function. Of course, 
this difference is merely technical but not principal, since both 
$\stackrel{\rightarrow}{r}$ and $\stackrel{\rightarrow}{h}$, according to 
inequalities (10) and (11), are slow variables.
 
The matrix $\hat A$ defined in Eq. (9) depends on time only through
$\stackrel{\rightarrow}{r}$ and $\stackrel{\rightarrow}{h}$. This permits 
us, introducing the small parameter
\begin{equation}
\label{12}
\varepsilon \equiv \sup\left \{ \frac{\omega_1}{\omega_2}\; ,
\frac{\omega}{\omega_2}\right \} \ll 1 \; ,
\end{equation}
to write the evolution equation
\begin{equation}
\label{13}
\frac{d\hat A}{dt} =\varepsilon \hat C \; ,
\end{equation}
in which $\hat C$ is a matrix with bounded elements for any $\varepsilon 
<1$ and all $t\geq 0$. Consequently, the matrix $\hat A$ has also to be 
treated as a slow matrix variable. This allows us to look for a 
particular solution of the spin equation (8) in the form
\begin{equation}
\label{14}
\stackrel{\rightarrow}{S}_i(t) = \stackrel{\rightarrow}{b}_i(t)
\exp\left\{ \varphi_i(t)\right\} \; ,
\end{equation}
in which $\stackrel{\rightarrow}{b}_i$, with $i=1,2,3$, are the eigenvectors
of the matrix $\hat A$, given by the eigenproblem 
$\hat A\stackrel{\rightarrow}{b}_i =\alpha_i\stackrel{\rightarrow}{b}_i$.
The nice antisymmetric structure of $\hat A$ makes it possible to easily 
organize that the set $\{\stackrel{\rightarrow}{b}_i\}$ would form an 
orthonormal basis. Substituting expression (14) into Eq. (8), we find
\begin{equation}
\label{15}
\varphi_i(t) = \int_0^t\left ( \omega_2\alpha_i -
\stackrel{\rightarrow}{b}_i^*\cdot \dot{\stackrel{\rightarrow}{b}}_i
\right ) dt \; ,
\end{equation}
as in Ref. [1]. As is clear, the form (14) is not, strictly speaking, an 
exact solution of Eq. (8), but this is what in the theory of differential 
equations called an {\it asymptotically exact} solution with respect to 
the small parameter $\varepsilon$, in the sense that this solution 
becomes exact when $\varepsilon\rightarrow 0$. The latter is evident from 
Eq. (13), since, when $\varepsilon \rightarrow 0$, the matrix $\hat A$ 
becomes a matrix with constant coefficients.

In the guiding center approach [27], which is a variant of the averaging
methods [21--25], the solution of type (14) is called the guiding center 
because it describes the leading behaviour of an exact solution. 
Corrections to the guiding center can be obtained by presenting the exact
solution as an expansion around this guiding center. Such expansions, 
because the guiding center itself is generally a function of the small 
parameter, are called the generalized asymptotic expansions. These have 
been introduced and studied by Lindstedt and Poincar\'e [28] and are 
actively used in averaging methods [24--26]. Corrections to the guiding 
center describe small fastly oscillating ripples of higher harmonics and 
with the amplitudes of the orders of increasing power of the small parameter.
As is well known [29--31], the construction of  generalized asymptotic 
series is not unique, being based on an incomplete expansion in powers of 
the small parameter. For instance, one can separately expand amplitudes 
and phases, as in the Lindstedt--Poincar\'e technique [28--31]. In such 
incomplete expansions, one may keep different number of terms in the 
amplitudes and phases, which is dictated by the properties of the 
particular equation. For example, this can be necessitated by cancelling 
secular terms or by considering more accurately the phase that, entering 
in an exponential, can influence the solution stronger than corrections 
to the amplitude. In our case, the nonuniqueness of defining the guiding 
center (14) could result in the possibility of writing, instead of the 
phase (15), the phase
\begin{equation}
\label{16}
\varphi_i(t) = \omega_2 \int_0^t \alpha_i(t) dt \; .
\end{equation}
This is admissible, since the second term in the integrand of Eq. (15) is 
small as compared to the first one. The sole doubt could be whether it is 
possible to omit the second term in Eq. (15) if one of the eigenvalues 
$\alpha_i$ is zero. In our case it is really so for $\alpha_3=0$. 
However, again owing to the nice antisymmetry property of $\hat A$, we 
can show that $\stackrel{\rightarrow}{b}_1^* =\stackrel{\rightarrow}{b}_2$
and $\stackrel{\rightarrow}{b}_3^* =\stackrel{\rightarrow}{b}_3$. Then, 
because of the normalization condition $|\stackrel{\rightarrow}{b}_3|^2 
=\stackrel{\rightarrow}{b}_3^2 = 1$, we have $2\stackrel{\rightarrow}{b}_3^*
\cdot\dot{\stackrel{\rightarrow}{b}}_3=d|\stackrel{\rightarrow}{b}_3|^2/dt=0$.
Thus, $\varphi_3(t)=0$ exactly. Which of the expressions, (15) or (16), is to
be used makes no difference for what follows, since at the next step of
the method [24--26] one has to substitute the solution for the fast variable
into the right--hand side of the equation for the slow variable and to 
average over time this right--hand side. In the course of this averaging, 
all fastly oscillating terms disappear, irrespectively to the fine structure
of their expressions.

When speaking about asymptotically exact solutions, one should not 
confuse this with the asymptotic behaviour with respect to small time
$t\rightarrow 0$. The solutions we are talking about are asymptotic with 
respect to small parameters, such as (10), (11), or (12). These parameters 
are always small for any $t\geq 0$. There is also the following invariance
property of Eq. (8) if $\hat A$ is an antisymmetric matrix. Let $[S_{ij}(t)]$
be the fundamental matrix of solutions to Eq. (8). Then [32], one has
$$
{\rm det} [ S_{ij}(t)] = {\rm det} [ S_{ij}(t_0)]
\exp\left\{ \omega_2\int_{t_0}^t {\rm Tr}\; \hat A(t')dt'\right \}
$$
for any $t_0$. For an antisymmetric matrix, ${\rm Tr}\hat A=0$. Hence, 
${\rm det}[ S_{ij}(t)]=const$. Thus, if $[S_{ij}(t_0)]$ is the 
fundamental matrix at some $t=t_0$, then the solutions 
$\stackrel{\rightarrow}{S}_i(t)$ are linearly independent for any $t\geq 
0$. The total solution of Eq. (8) is the linear combination
\begin{equation}
\label{17}
\stackrel{\rightarrow}{S}(t) = \sum_{i=1}^3 
a_i\stackrel{\rightarrow}{S}_i(t)
\end{equation}
of linear independent particular solutions, with the coefficients 
$a_i=\stackrel{\rightarrow}{S}_0\cdot\stackrel{\rightarrow}{b}_i(0)$
defined by initial conditions. In our case, the asymptotically exact 
particular solutions are given by Eq. (14).

Let us emphasize that all the consideration is based on and justified 
by the averaging methods [21--26] requiring the existence of small 
parameters that permit one to classify functional variables onto fast
and slow. Without such small parameters we would have quite a different
story.

After the solution (17) for the fast variable is found, then following 
the multifrequency averaging methods [24--26], one has to substitute it 
into the equation (7) for the slow variable and to average over time this 
right--hand side, which gives
\begin{equation}
\label{18}
\frac{d^2\stackrel{\rightarrow}{r}}{dt^2} =
\stackrel{\rightarrow}{F} \equiv \omega_1^2 < S_x\stackrel{\rightarrow}{e}_x
+ S_y\stackrel{\rightarrow}{e}_y + 
\lambda S_z\stackrel{\rightarrow}{e}_z > \; .
\end{equation}
Till now, the concrete form of the transverse field (4) has been of no 
importance, provided that condition (11) holds true. To explicitly 
accomplish the time averaging in Eq. (18), we need to specify the field 
$\stackrel{\rightarrow}{h}$. With the rotating field (5), we find
$$
\stackrel{\rightarrow}{F} = \frac{1}{2} f_1 \omega_1^2 \left [ (1 +x ) S_x^0
+ y S_y^0 + \lambda z S_z^0 \right ] \left (
x\stackrel{\rightarrow}{e}_x + y\stackrel{\rightarrow}{e}_y +
2\lambda^2 z \stackrel{\rightarrow}{e}_z \right ) \; ,
$$
where
$$
f_1 \equiv \left [ ( 1 + 2x + x^2 + y^2 + \lambda^2 z^2)
( 1 + x^2 + y^2 + \lambda^2 z^2 ) \right ]^{-1/2} .
$$

As is shown in paper [16], the semiconfining regime of motion can be 
realized taking the initial spin polarization given by the initial conditions
\begin{equation}
\label{19}
S_x^0 = S_y^0 = 0 \; , \qquad S_z^0 \equiv S \neq 0 \; .
\end{equation}
Such initial conditions can be prepared in several ways. The first 
possibility is to confine atoms in a trap, in which all atoms are 
polarized having their spins in the $z$ direction, which e.g. can be done 
by means of the trap of Ref. [33], being a quadrupole trap with a bias 
field along the $z$ axis. Then the longitudinal bias field is quickly 
switched off and, at the same time, a transverse field, as in Ref. [34], 
is switched on. Thus, we get the desired initial conditions. Another 
possibility is to prepare spin polarized atoms in one trap and quickly 
load them into another trap with the required field configuration. The 
possibility of realizing two ways of transferring atoms from one trap to 
another, by means of sudden transfer as opposed to adiabatic transfer, is 
discussed in Ref. [35]. The third possibility could be by acting on the 
trapped atoms with a short pulse of strong magnetic field, polarizing 
atomic spins in the necessary way.

With the initial conditions (19), equation (18) reduces to the system
\begin{equation}
\label{20}
\frac{d^2x}{dt^2} = \frac{\lambda}{2} S\omega_1^2 f_1 z x\; ,
\qquad
\frac{d^2z}{dt^2} = \lambda^3 S \omega_1^2 f_1 z^2 \; ,
\end{equation}
the equation for the $y$ component having the same form as for the $x$ 
component. The system of equations (20) can be solved analytically only 
for the case $|\stackrel{\rightarrow}{r}|\ll 1$ when $f_1\rightarrow 1$, 
as is considered in Ref. [16]. This consideration proves the appearance of 
the semiconfining regime of motion and the existence of collimation with 
the aspect ratio
$$
\frac{x(t)}{z(t)} \sim | t - t_0 |^{3/2} ,
$$
as $t\rightarrow t_0$, where $t_0$ is defined by the initial conditions 
for the space variable. When $|\stackrel{\rightarrow}{r}|$ becomes not 
small, one has to resort to numerical solution which has been done in 
Refs. [18,20] confirming the existence of the semiconfining regime and the 
formation of well-collimated atomic beams.

Let us stress again that asymptotic solutions obtained by the averaging 
methods are asymptotic with respect to small parameters and have nothing 
to do with short--time approximation. As follows from the theory of the 
averaging methods [21--25], the latter provide accurate solutions to 
differential equations for long time intervals.

\section{Applicability to Different Fields}

The semiconfining regime of motion can be realized for a wide class of 
magnetic fields. Not only the anisotropy parameter $\lambda$ in the 
quadrupole field (3) can be varied, but also different transverse fields 
(4) can be employed. In the previous section, the particular form of a 
transverse field has not been specified till formula (18), showing by 
this that the same consideration is valid for any transverse field 
satisfying condition (11). Now we shall show that the choice of the 
rotating field (5) is not compulsory and other transverse fields can be 
taken provided condition (11) holds true.

Let us consider the simple case of a constant transverse field
\begin{equation}
\label{21}
h_x = const \; , \qquad h_y = const \; ,
\end{equation}
where, similarly to Eq. (4), it is assumed that 
$|\stackrel{\rightarrow}{h}|=1$. Then the small parameter (11) is 
identically zero.

With the transverse field (21), the right--hand side of Eq. (18) becomes
\begin{equation}
\label{22}
\stackrel{\rightarrow}{F} = f_2\omega_1^2 \left [ (x + h_x) S_x^0 +
(y + h_y) S_y^0 + \lambda z S_z^0 \right ]
\left [ (x + h_x)\stackrel{\rightarrow}{e}_x +
(y + h_y)\stackrel{\rightarrow}{e}_y +
\lambda^2 z\stackrel{\rightarrow}{e}_z \right ] \; ,
\end{equation}
where
$$
f_2 \equiv 
\left [ ( x + h_x )^2 + ( y + h_y)^2 + \lambda^2 z^2 \right ]^{-1} . 
$$
Accepting the initial conditions (19), we come to the equations
\begin{equation}
\label{23}
\frac{d^2x}{dt^2} = \lambda S\omega_1^2 f_2 z ( x + h_x ) \; ,
\qquad
\frac{d^2z}{dt^2} = \lambda^3 S \omega_1^2 f_2 z^2 \; .
\end{equation}
One may notice that these equations are invariant with respect to the 
change $\lambda\rightarrow -\lambda$, $S\rightarrow -S$ or
to the change $S\rightarrow -S$, $z\rightarrow -z$. Therefore, it would 
be sufficient to consider a case of one fixed sign for $\lambda S$, say, 
$\lambda S>0$. Passing to another sign of $\lambda S<0$ is equivalent to 
the inversion $z\rightarrow -z$.

Equations (23) are very similar to Eqs. (20). In the same way as in Ref. 
[16], we can demonstrate that the solution to Eqs. (23) corresponds to the 
semiconfining regime of motion. For $|\stackrel{\rightarrow}{r}|\ll 1$, when 
$f_2\rightarrow 1$, we have
\begin{equation}
\label{24}
\frac{d^2x}{dt^2} = \lambda S\omega_1^2 h_x z \; , \qquad
\frac{d^2z}{dt^2} = \lambda^3 S\omega_1^2 z^2 .
\end{equation}
The second equation here is the same as in Eq. (20) if $f_1\rightarrow 
1$. The analysis shows that the motion along the $z$ axis is 
semiconfined. The collimation of the forming directed beam is 
characterized by the aspect ratio
\begin{equation}
\label{25}
\frac{x(t)}{z(t)} \sim | t - t_0|^2 \ln | t - t_0 | \; ,
\end{equation}
as $t\rightarrow t_0$. Comparing this with the aspect ratio for the rotating 
field, we take into account that for $\tau \equiv |t-t_0|\ll 1$ one has 
$\tau^{1/2}|\ln\tau|\ll 1$. Therefore, the constant transverse field 
provides even better collimation than the rotating field.

\section{Influence of Gravitational Force}

\vspace{3mm}

Atoms having mass are certainly subject to the action of the gravitational 
force. For the adiabatic motion of atoms confined in a trap, this force 
is not as significant since it can always be easily compensated by an 
additional gradient magnetic field. But what is the role of this force in 
a nonadiabatic motion? The principal answer to this question  is, 
actually, almost evident: If the gravitational force can be compensated 
by additional gradient fields, it will lead to no principal changes in 
the motion of atoms inside a trap. However, to be precise, let us turn to
mathematics.

Recall first of all that the coordinate axes everywhere in our formulas 
have been related to the magnetic field configuration, so that the axis 
$z$ is an axis of the device, but not necessary the vertical axis related 
to gravity. As far as the device can be oriented arbitrarily, the 
gravitational force can also be directed along different axes. For 
instance, assume that this force is $-mg\stackrel{\rightarrow}{e}_x$, 
where $g\approx 10^3$ cm s$^{-2}$ is the standard gravitational 
acceleration. Adding this force to the right--hand side of Eq. (18) 
results, instead of Eqs. (20), in the equations
\begin{equation}
\label{26}
\frac{d^2x}{dt^2} = \omega_1^2\left (\frac{\lambda}{2} S f_1 z x -
\alpha \right ) \; ,
\qquad
\frac{d^2z}{dt^2} = \lambda^3 S \omega_1^2 f_1 z^2 \; ,
\end{equation}
in which the notation
\begin{equation}
\label{27}
\alpha \equiv \frac{g}{R_0\omega_1^2}
\end{equation}
is used. The occurrence of an additional term in the first of equations 
(26) will lead to the gravitational drift along the $x$ axis. To 
compensate the gravitational force, an auxiliary gradient magnetic field 
is to be superimposed. This is equivalent to the increase of the 
coefficient near $\stackrel{\rightarrow}{e}_x$ in the quadrupole field 
(3), which yields the scaling $x\rightarrow \lambda_x x$ in Eq. (26). One 
may also increase the anisotropy coefficient $\lambda$. The final aim is 
to make the ratio $\alpha/\lambda\lambda_x$ small, which thus reduces the 
influence of the gravity force. This ratio is actually small already for 
existing traps, without involving additional compensating fields. For 
instance, taking the values $\omega_1\sim(10^2 - 10^3)$s$^{-1}$, 
$R_0\sim(0.1 - 0.5)$cm, $|\lambda|=2$, and $\lambda_x=1$, typical of
many modern traps, we have $\alpha/\lambda\lambda_x$ of the order 
$10^{-1} -10^{-3}$. Increasing the parameters $\lambda$ and $\lambda_x$, 
by switching on additional gradient fields, the ratio
$\alpha/\lambda\lambda_x$ can be made arbitrarily small, thus making the 
influence of the gravitational force negligible.

Moreover, we have an additional possibility of changing the orientation 
of the device with respect to the gravitational force. For instance, the 
latter can be directed along the $z$ axis, so that this force is 
$-mg\stackrel{\rightarrow}{e}_z$. Then, instead of Eqs. (26), we get
\begin{equation}
\label{28}
\frac{d^2x}{dt^2} = \frac{\lambda}{2} S \omega_1^2 f_1 x z \; ,
\qquad
\frac{d^2z}{dt^2} = \omega_1^2 ( \lambda^3 S f_1 z^2 - \alpha ) \; .
\end{equation}
From here it is seen that the influence of the gravitational force can be 
reduced by increasing the anisotropy parameter $\lambda$. To demonstrate 
this more accurately, let us consider the same case as earlier, when
$|\stackrel{\rightarrow}{r}|\ll 1$. Then, integrating once the second 
equation in the system (28), we have
$$
\left ( \frac{dz}{dt}\right )^2 = \frac{2}{3} \lambda^3 S \omega_1^2
\left [ z^3 - z_0^3 - \frac{3\alpha}{\lambda^3 S} ( z - z_0 )
+\zeta \right ] \; ,
$$
where
$$
\zeta \equiv \frac{3\dot{z}_0^2}{2\lambda^3 S\omega_1^2} \; .
$$
The latter equation, by means of the substitution
\begin{equation}
\label{29}
z(t) = \frac{6}{\lambda^3 S} {\cal P}(\tau -\tau_0) \; , 
\qquad \tau\equiv \omega_1 t \; ,
\end{equation}
can be transformed to the Weierstrass equation
$$
\left ( \frac{d{\cal P}}{d\tau} \right )^2 = 4{\cal P}^3 - g_2{\cal P} - g_3 
\; , $$
with the Weierstrass invariants
$$
g_2= \frac{\alpha}{3}\;\lambda^3 S \; , \qquad
g_3 = \frac{(\lambda^3 S)^3}{54} \; \left ( 
z_0^3 - \frac{3\alpha}{\lambda^3 S} \; z_0 -\zeta\right ) \; .
$$
From the properties of the Weierstrass function ${\cal P}(\tau)$ it 
follows that the semiconfining regime of motion is realized when 
$g_2^3 < 27g_3^2$, which gives
\begin{equation}
\label{30}
\frac{4\alpha^3}{\lambda^9 S^3}\; <\;  \left (
z_0^3 - \frac{3\alpha}{\lambda^3 S} z_0 - \zeta \right )^2 \; .
\end{equation}
As is evident form this inequality, the terms related to the gravity can 
be strongly reduced by increasing the anisotropy parameter $\lambda$, 
that is by increasing the field gradient in the $z$ direction. Even for 
the typical value $|\lambda|=2$, without involving additional 
compensating fields, the left--hand side of the inequality (30) is of the 
order $10^{-3}\alpha^3$. For $\alpha\sim 10^{-3}$ this yields $10^{-12}$, 
which is very small. Increasing the parameter $\lambda$ can practically 
completely eliminate the influence of gravity. Note also that for the 
case $\lambda S<0$, the inequality (30) is always valid since its 
left--hand side is negative while the right--hand side is nonnegative. 
The latter case shows that the gravitational force even may help to 
realize semiconfinement, which happens when the direction of this force 
coincides with that of the atomic escape. In this way, it is not difficult 
to prepare such magnetic fields and to choose the orientation of the device 
so that the gravitational force would not essentially disturb the 
semiconfining regime of motion.

\section{Role of Random Collisions}

In order to investigate the role of atomic collisions, we have to 
compliment the right--hand side of the evolution equation (18) by an 
additional term describing these collisions, so that we can write
\begin{equation}
\label{31}
\frac{d^2\stackrel{\rightarrow}{r}}{dt^2} =
\stackrel{\rightarrow}{F} + \gamma\stackrel{\rightarrow}{\xi} \; ,
\end{equation}
where $\stackrel{\rightarrow}{F}$ is the same expression as in Eq. (18), 
$\gamma$ is a collision rate, and $\stackrel{\rightarrow}{\xi}$ is a 
vector whose properties are defined by the details of the pair 
collisions. Two principally different situations can exist. One is when 
the effective collision force $\gamma\stackrel{\rightarrow}{\xi}$ is larger
than or comparable with $\stackrel{\rightarrow}{F}$. Then it is clear 
that the atomic motion is essentially governed by the collision force 
whose particular properties become important. However, this case is of no 
interest for us since, as is evident, in the presence of strong and 
frequent collisions no directed semiconfined motion can occur. Any 
organized directed motion will be strongly suppressed by disorganizing 
random collisions. The second situation is when 
$\gamma\stackrel{\rightarrow}{\xi}$ is much weaker than 
$\stackrel{\rightarrow}{F}$. And solely this case is of interest since 
only then an organized unidirectional motion can arise. But when the 
force $\gamma\stackrel{\rightarrow}{\xi}$ representing random pair 
collisions is just a weak perturbation, then the details of this force 
are not of great importance and it can be modelled by the stochastic 
white--noise variable. It is this approach that was employed in Ref. [16], 
where some additional simplifications for $\stackrel{\rightarrow}{\xi}$ 
were assumed. These simplifications are not principal, and the 
consideration can be easily generalized to the case of an anisotropic 
random vector $\stackrel{\rightarrow}{\xi} =\{\xi_\mu\}$, where $\mu=x,y,z$.
In the general anisotropic case, the stochastic properties of the set 
$\{\xi_\mu\}$ of white--noise random variables is characterized by the 
stochastic averages
\begin{equation}
\label{32}
\ll \xi_\mu(t) \gg = 0 \quad (\mu=x,y,z) \; ,
\qquad
\ll \xi_\mu(t)\xi_\nu(t') \gg = 2 D_\mu\delta_{\mu\nu}\delta(t-t') \; ,
\end{equation}
in which $D_\mu$ is a diffusion rate in the $\mu$ direction.

Adding the anisotropic random force to Eq. (20), for 
$|\stackrel{\rightarrow}{r}|\ll 1$, we have
\begin{equation}
\label{33}
\frac{d^2x}{dt^2} = \frac{\lambda}{2} S\omega_1^2 z x +\gamma\xi_x \; ,
\qquad
\frac{d^2z}{dt^2} = \lambda^3 S\omega_1^2 z^2 +\gamma\xi_z \; .
\end{equation}
As far as only the case of small disturbance of the semiconfining regime 
of motion is of our concern, we can solve Eqs. (33) by perturbation 
theory writing the solutions as
\begin{equation}
\label{34}
x=x_1 + x_2 \; , \qquad z = z_1 + z_2 \; ,
\end{equation}
where $x_1$ and $z_1$ are the solutions of the unperturbed equations (20) 
and $x_2$ and $z_2$ are the solutions to the equations
\begin{equation}
\label{35}
\frac{d^2x_2}{dt^2} =\frac{\lambda}{2} S \omega_1^2 ( z_1x_2 + x_1 z_2 ) +
\gamma\xi_x \; ,
\qquad
\frac{d^2z_2}{dt^2} = 2\lambda^3 S \omega_1^2 z_1 z_2 +\gamma\xi_z \; .
\end{equation}
From here, similarly to the way of Ref. [16], we get
$$
x_2(t) =\int_0^t G_x(t-t') \left [ \gamma\xi_x(t') +
\frac{\lambda}{2} S\omega_1^2 x_1 z_2(t')\right ] dt' \; ,
$$
\begin{equation}
\label{36}
z_2(t) = \int_0^t G_z(t-t')\gamma\xi_z(t') dt' \; ,
\end{equation}
where
$$
G_x(t) =\frac{{\rm sinh}(\beta t)}{\beta} \; , \qquad
G_z(t) = \frac{{\rm sinh}(\beta_z t)}{\beta_z} \; , \qquad
\beta =\left ( \frac{\lambda}{2} S \omega_1^2 z_1\right )^{1/2} ,
\qquad \beta_z = 2|\lambda|\beta \; .
$$
Because of Eqs. (32), we have $\ll x_2 \gg\; =\;\ll z_2 \gg\; =0$. And for 
the mean square deviations, we obtain
$$
\ll x^2 \gg\; = \frac{\gamma^2D_xt}{\beta^2} \left [
\frac{{\rm sinh}(2\beta t)}{2\beta t} - 1 \right ] +
\frac{\beta^4x_1^2\gamma^2D_zt}{\beta_z^2(\beta_z^2-\beta^2)^2z_1^2} \times
$$
\begin{equation}
\label{37}
\times
\left\{ \frac{{\rm sinh}(\beta_zt)}{\beta_zt} \left [
{\rm cosh}(\beta_zt) - {\rm cosh}(\beta t)\right ] -
\frac{\beta_z}{\beta}{\rm sinh}(\beta_zt){\rm sinh}(\beta t) + 
{\rm cosh}(\beta_zt){\rm cosh}(\beta t) - 1 \right \} \; ,
\end{equation}
\begin{equation}
\label{38}
\ll z^2 \gg\; =\frac{\gamma^2D_zt}{\beta_z^2}\left [
\frac{{\rm sinh}(2\beta_zt)}{2\beta_zt} - 1 \right ] \; .
\end{equation}
These formulas generalize the result of Ref. [16]. However the main aim in 
considering the role of random collisions is not just to derive formulas 
as above but rather to find conditions under which these collisions would 
not essentially disturb the semiconfined motion. Such a condition can be 
written as
\begin{equation}
\label{39}
\frac{\gamma^2D}{\omega_1^3} \ll 1 \; , \qquad
D \equiv \sup \{ D_x,D_y,D_z\} \; .
\end{equation}
For estimates, we can take the collision rate as $\gamma\sim \hbar\rho 
a_s/m$, where $\rho$ is the density of atoms and $a_s$, scattering length, 
and for the diffusion rate we may write $D\sim k_BT/\hbar$, where $T$ is 
temperature. Then condition (39) yields
\begin{equation}
\label{40}
(\rho a_s^3)^2 k_BT\left (\frac{\hbar^2}{ma_s^2}\right )^2 \ll
(\hbar\omega_1)^3 \; .
\end{equation}
This shows that random atomic collisions will not disturb much the 
organized semiconfined motion if density and temperature are small enough 
to satisfy condition (40).

\vspace{5mm}

{\bf Acknoledgements}

\vskip 2mm

We are grateful for discussions to V.S. Bagnato. Financial support from the
S\~ao Paulo State Research Foundation is appreciated.

\end{document}